\documentclass[12pt]{article}
\usepackage{lipsum}
\usepackage{amsmath} 
\usepackage{amssymb}
\usepackage{hhline}
\usepackage[scale={.75,.75}]{geometry}
\usepackage{url}
\usepackage{lscape}
\usepackage{caption}
\usepackage[numbers,sort&compress]{natbib}
\usepackage{epsfig}
\usepackage{multirow} 
\usepackage{color}
\usepackage{verbatim}
\usepackage{nicefrac}
\usepackage{upgreek}
\usepackage{bbm}
\usepackage{setspace}
\usepackage{pstricks}
\usepackage{psfrag}
\usepackage{color}

\newcommand{\slashed}[1]{\displaystyle{\not}{#1}}

\usepackage{hyperref}% add hypertext capabilities
\definecolor{green}{rgb}{0,0.5,0}

\begin{document}
\date{}

\title{\vspace{-2.5cm} 
\begin{flushright}
\vspace{-0.4cm}
{\scriptsize \tt TUM-HEP-981/15}  % 11-18
\end{flushright}
{\bf Theoretical Status of Neutrino Physics
}
}

\author{Marco Drewes\\ 
\footnotesize{Physik Department T70, Technische Universit\"at M\"unchen, }\\
\footnotesize{James Franck Stra\ss e 1, D-85748 Garching, Germany}}

\maketitle

\begin{abstract}
\noindent %I summarise my viewpoint on the theoretical status of neutrino physics. The key question remains what is the origin of neutrino masses.
In the framework of renormalisable relativistic quantum field theory,
the explanation of neutrino masses necessarily requires the existence of new physical states. These new states may also be responsible for other unexplained phenomena in particle physics and cosmology. After a brief introduction, I focus on scenarios in which the neutrino masses are generated by the type-I seesaw mechanism and review the phenomenological implications of different choices of the seesaw scale.
\end{abstract}
%\vspace{8cm}
\noindent This mini-review is based on my talk at the 16th International Workshop on Neutrino Factories and Future Neutrino Beam Facilities (NUFACT2014),  25-30 August 2014 at the University of Glasgow (United Kingdom). 
In order to increase the usefulness of this document, I decided to add a number of figures with caption that do not appear in the proceedings due to length restrictions. 
In spite of the ambitious title, neither my talk nor the present summary come even close to a complete discussion of the topic. I would like to apologise to all authors whose important contributions I have left out.
\tableofcontents
\newpage

%\tableofcontents

\section{Why are neutrinos so interesting?}
\begin{figure}
  \centering
    \includegraphics[width=7cm]{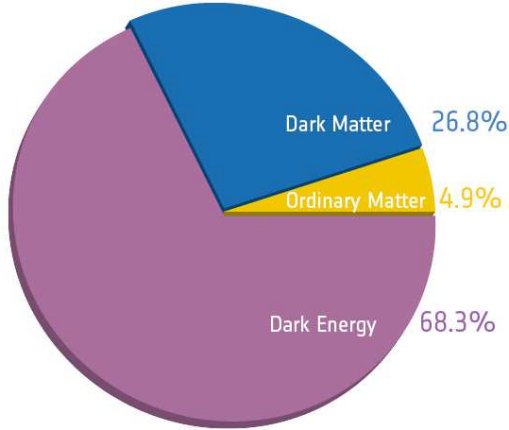}
    \caption{Neutrino oscillations (1) are the only established proof of physics beyond the SM that has been found in the laboratory. The puzzles (2)-(4) are all inferred from observations in outer space. A particularly useful probe is the CMB, which (with some input from other sources) allows to precisely measure the amount of ordinary baryonic matter (2) and DM (3) in the universe, shown here as fractions of the total energy density of the observable universe.
Diagram taken from \emph{planck.cf.ac.uk}.\label{CosmicPie}
}
\end{figure}

The Standard Model of particle physics and theory of general relativity form the basic pillars of modern physics.
While there are many emergent phenomena in nature the understanding of which poses a great challenge due to their complexity, only four observations have been confirmed\footnote{Sometimes the acceleration of the cosmic expansion at present time is included in this list as "dark energy". However, all present observational data is consistent with a cosmological constant, which is simply a free parameter in general relativity.} 
that cannot in principle be understood this framework \cite{Agashe:2014kda}:
\footnote{In addition to this empirical evidence, there are a number of aesthetic issues that appear unsatisfactory from a theory viewpoint, such as the hierarchy problem, strong CP-problem, the flavour puzzle and the value of the cosmological constant. Moreover, it is not clear what is the correct description of quantum gravity, Though extremely interesting, the latter is not required to explain any experiment in foreseeable time.} 
\begin{itemize}
\vspace{-0.3cm}
\item[(1)] \emph{neutrino flavour oscillations}, 
\vspace{-0.3cm}
\item[(2)] \emph{the baryon asymmetry of the universe (BAU)}, i.e. the tiny excess of matter over antimatter in the early universe, which explains the presence of matter in the cosmos at present time as the ``leftover'' after mutual annihilation of all antimatter with matter,\footnote{See \cite{Canetti:2012zc} for a recent review of the evidence for this interpretation.} 
\vspace{-0.3cm}
\item[(3)] the \emph{composition and origin of the Dark Matter (DM)} that appears to make up most of the mass of galaxies and galaxy clusters and 
\vspace{-0.3cm}
\item[(4)] the \emph{hot big bang} initial conditions, in particular the overall homogeneity and isotropy of the early universe seen in the cosmic microwave background (flatness and horizon problems).\footnote{The idea of \emph{cosmic inflation} \cite{Starobinsky:1980te} provides an elegant solution to this problem, but it is not known what mechanism drove the accelerated expansion.}
\end{itemize}
%\vspace{-0.3cm}
Neutrino oscillations (1) are the only one amongst these that have been observed in the laboratory, see Fig.~\ref{CosmicPie}. 
It is sometimes argued that a mass term that would explain can be added "trivially" within the SM. It should, however, be clear that the construction of such a mass term in the framework of renormalisable quantum field theory necessarily requires the introduction of new physical states. 

In the SM all fermion masses are Dirac masses, which are generated via the Higgs mechanism from Yukawa couplings, e.g. $\overline{e_{L}}m_e e_{R}$ for charged leptons.
In order to write down such a term for neutrinos, one necessarily has to add right handed neutrinos $\nu_R$ to the SM Lagrangian $\mathcal{L}_{SM}$. If a Dirac mass $\overline{\nu_{L}}m_D\nu_{R}$ is the only source of neutrino masses, then the mass generation is exactly the same as for charged leptons and quarks. Then neutrinos are Dirac particles and the lepton sector resembles the quark sector without strong interaction (and with a different choice of the numerical parameters). 
Then $\nu_L$ and $\nu_R$ form a Dirac spinor $\Psi_\nu=\nu_L+\nu_R$, and the new degrees of freedom $\nu_R$ would not appear as "new particles", but rather lead to additional spin states for the neutrinos. 
However, as gauge singlets the $\nu_R$ can have a Majorana mass term $\overline{\nu_R} M_M \nu_R^c$.
$M_M$ in general is not diagonal in the flavour basis where $m_D$ is diagonal. Diagonalising the full mass term leads to Majorana neutrinos and new mass eigenstates. These \emph{sterile neutrinos} appear as new physical particles that can be responsible for various phenomena, including (2) and (3), see \cite{Drewes:2013gca} for a recent review. If one wants to avoid this, one has to forbid $M_M$ by postulating an additional symmetry (e.g.\ lepton number conservation).

\begin{figure}
\begin{center}
%\vspace{-2cm}
%\uncover<2->{
\includegraphics[width=150mm]{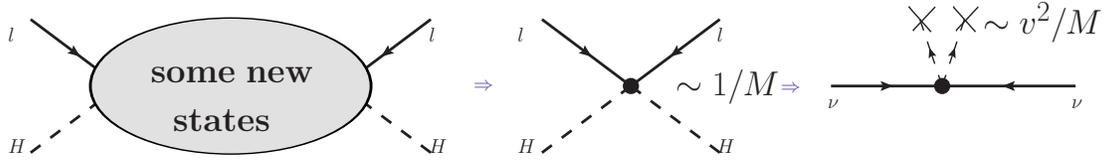}
%}
\end{center}
\caption{\emph{First step}: 
Symbolic representation of the generation of the dimension five operator (\ref{Weinberg}) by ``integrating out'' some unknown new state(s) with mass $\sim M\gg v$.
If the states are very heavy, they do not propagate as real particles in processes at energies $\ll M$.
Then the Feynman diagram on the left can in good approximation be replaced by the local ``contact interaction'' vertex represented by the black dot in the diagram in the middle.
This is in analogy to the way how the four fermion interaction $\propto G_F$ is obtained from integrating out the weak gauge bosons.
\emph{Second step}:
At energies far below the Higgs mass, no Higgs particles can be produced, and the only appearance that the Higgs field makes is via its vev $v$. By replacing $H\rightarrow(0,v)^T$ the Weinberg operator (\ref{Weinberg}) is turned into a bilinear in the neutrino fields $\frac{v^2}{2}\bar{\nu}_LfM^{-1}\nu_L$ that acts as Majorana mass term with $m_\nu=v^2fM^{-1}$ and can be represented by the diagram on the right. Here the dashed Higgs lines that end in a cross represent the insertion of a Higgs vev $v$. They and the effective vertex are often omitted, so that the Majorana mass term is simply represented by ``clashing arrows''.
The order of the two steps (in terms of energy scales) can be the other way around if $M<v$, but the result at energies much smaller than $v$ and $M$ is the same.\label{WeinbergFig} 
}
\end{figure}

One could attempt to directly write down a Majorana mass $\overline{\nu_L} m_\nu \nu_L^c$ term for  the LH neutrinos $\nu_L$, which in this case are Majorana particles. 
While gauge invariance forbids such a term at the "fundamental" level, it can be generated via the Higgs mechanism form the Weinberg operator \cite{Weinberg:1979sa}, see Fig.~\ref{WeinbergFig}
\begin{equation}\label{Weinberg}
\frac{1}{2}\bar{L_{L}}\tilde{H}\frac{f}{M}\tilde{H}^{T}L_{L}^{c}
\end{equation}  
with $m_\nu=v^2f M^{-1}$, where $f$ is a flavour matrix, $v$ is the Higgs vev 
and $M$ characterises mass scales far above the energy of neutrino experiments.
Indeed all experimental neutrino data can be explained by adding this operator to the SM. However, the new term is not renormalisable. In a fundamental theory it should be generated by "integrating out" some heavier new states with masses $\sim M$. 
This could be RH neutrinos $\nu_R$ with a Majorana mass $M_M$ (in which case $M\sim M_M$), see section \ref{part:seesaw}, but there are numerous other possibilities, see e.g. \cite{Mohapatra:1998rq} for a summary. Exploring these is the goal of neutrino model building.
The physical neutrino mass squares $m_i^2$ are given by the eigenvalues of $m_\nu m_\nu^\dagger$ and can conveniently be read off as $|m_i|$ after diagonalising $m_\nu=U_\nu{\rm diag}(m_1,m_2,m_3)U_\nu^T$, where with $U_\nu$ is the neutrino mixing matrix in the charged lepton mass basis. 
The past few years have seen enormous progress in determining the parameters in $m_\nu$. All three mixing angles and two mass splittings $|m_i^2-m_j^2|$ have been measured, and experimental data may soon allow to constrain the CP-violating Dirac phase  \cite{Gonzalez-Garcia:2014bfa} and the mass ordering \cite{Blennow:2013oma}.

In contrast, next to nothing is known about the new states that generate $m_\nu$. Though the measurement of the mass splittings and mixings angles is an immense experimental achievement, and the finding of CP-violation in $m_\nu$ would clearly be a milestone achievement in experimental physics, the ultimate goal remains to identify the new states and unveil the mechanism of neutrino mass generation.
There are two possible explanations why they have not been found to date. Either they are much heavier than the W-boson (``energy frontier''), or they have very feeble interactions (``intensity frontier''). Of course, a discovery is only possible if the new mass scale $M$ is within reach of experiments, i.e.\ at the TeV scale or below, and even then the search might be very challenging. However, a discovery would not only clarify the origin of neutrino masses, but the physics behind their generation may also be responsible for other phenomena including (2) and (3).
Hence, neutrino masses 
may act as a "portal" to a (possibly more complicated) unknown/hidden sector and yield the answer to deep questions in cosmology, such as the origin of matter and dark matter.

\section{Neutrino masses}
Any model of neutrino masses should address the fact that they are orders of magnitude smaller than any other fermion masses in the SM ("mass puzzle"). It is very convenient to classify models according to the way how this hierarchy is explained. \\
\begin{itemize}
\item 
\textbf{Small coupling constant}: A tiny coupling constant can explain the smallness of $m_i$ generated via spontaneous symmetry breaking, but it would have to be very tiny. For instance, Dirac masses generated via the standard Higgs mechanism would require $F\sim 10^{-12}$, which is considered ``unnatural'' by most theorists.\\  
\item 
\textbf{Seesaw mechanism}: If the $m_i$ are generated at classical level, they may be suppressed by the new heavy scale $M$. The most studied version is the type-I seesaw \cite{Minkowski:1977sc} discussed in section \ref{part:seesaw}, the two other possibilities \cite{Ma:1998dn} are the type-II \cite{Cheng:1980qt}
and type-III \cite{Foot:1988aq} seesaw.\\
\item 
\textbf{Flavour ("horizontal") symmetry}: Individual entries of the matrix $m_\nu$ may be large, but still yield small $m_i$ if there is a symmetry that leads to approximate lepton number conservation and cancellations in $m_\nu m_\nu^\dagger$. 
Prominent examples of this class are Froggatt-Nielsen type models \cite{Froggatt:1995rt}, the inverse seesaw \cite{Mohapatra:1986bd} and other models with approximate lepton number conservation, e.g. \cite{Chikashige:1980ui}.\\
\item 
\textbf{Radiative masses}: Neutrinos could be classically massless and their masses generated by quantum corrections. The suppression by the "loop factor" $(4\pi)^2$ is not sufficient to explain the smallness of the $m_i$, but the interaction of the new particles in the loop with $\nu_L$ may involve some small coupling constants that do the job. Flavour symmetries or an additional seesaw-like suppression can help to make such models more "natural", see e.g. \cite{Zee:1980ai,Khoze:2013oga}.
\end{itemize}
Of course, any combination of these ideas could be realised in nature.
In addition to the smallness of the mass eigenvalues, it would be desirable to find an explanation for the observed flavour structure of $m_\nu$ ("flavour puzzle"). 
While the quark mass matrix shows a distinct structure (being approximately diagonal in the weak interaction basis), there is no obvious symmetry in $m_\nu$. Numerous attempts have been made to impose more subtle discrete or continuous symmetries, see e.g. \cite{King:2014nza}.
The basic problem is that the reservoir of possible symmetries is practically unlimited, so for any possible observed $m_\nu$ one could find some kind of symmetry that "predicts" it. Since we already know a lot about the flavour structure, models can only be convincing if they either predict other (yet unmeasured) observables or are "simple" and esthetically convincing from some viewpoint.
Prior to the measurement of $\theta_{13}$ there seemed to be at least some kind of structure, and models predicting $\theta_{13}=0$ seemed well-motivated, such as tri/bi-maximal mixing \cite{Harrison:1999cf}.
With the present data, it seems very difficult to single out any class of models that could explain $m_\nu$ in terms of a simple symmetry and/or a small number of parameters, and interest in anarchic models \cite{Hall:1999sn} with random values has grown.

\begin{figure}
  \centering
    \includegraphics[width=9cm]{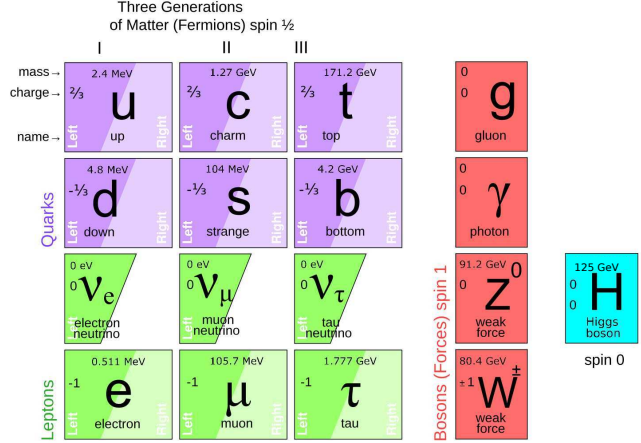}
    \caption{The particle content of the SM. Are we missing the right handed partners of the neutrinos? Picture taken from \cite{Shaposhnikov:2013dra}.\label{SM1}}
\end{figure}

In the following I focus on those scenarios in which RH neutrinos $\nu_R$ generate $m_\nu$ via the seesaw mechanism. While the seesaw mechanism alone is unable to predict the flavour structure, it at least explains the smallness of neutrino masses. The existence of $\nu_R$ seems very well-motivated because all other known fermions exist with both, LH and RH chirality, see Fig.~\ref{SM1}. Moreover, $\nu_R$ appear in 
many popular theories (left-right symmetric models,  SO(10) grand unified theories, generally all theories with a $U(1)_{B-L}$ symmetry) and can be related to various other phenomena in cosmology and particle physics, such as dark matter, baryogenesis, dark radiation and neutrino oscillation anomalies, see Appendix \ref{overview} for an overview and Refs.~\cite{Abazajian:2012ys,Drewes:2013gca} for more detailed reviews. 
 
\section{Probing the Seesaw Mechanism}\label{part:seesaw}
\begin{figure}
  \centering
    \includegraphics[width=9cm]{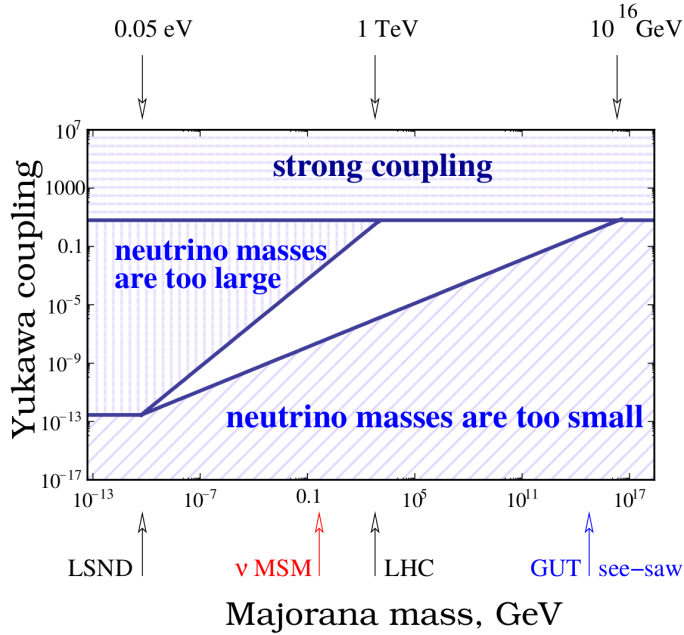}
    \caption{A schematic illustration of the relation between ${rm tr}F^\dagger F$ and $M_I$ imposed by neutrino oscillation data (plot taken from Ref.~\cite{Abazajian:2012ys}). Individual elements of  $F$ can deviate considerably from $F_0$ if there are cancellations in (\ref{activemass}). Cosmological constraints allow to further restrict the mass range, see Fig.~\ref{allowedrange}\label{seesawfig}}
\end{figure}

\begin{figure}
\begin{center}
%\vspace{-2cm}
%\uncover<2->{
\includegraphics[width=65mm]{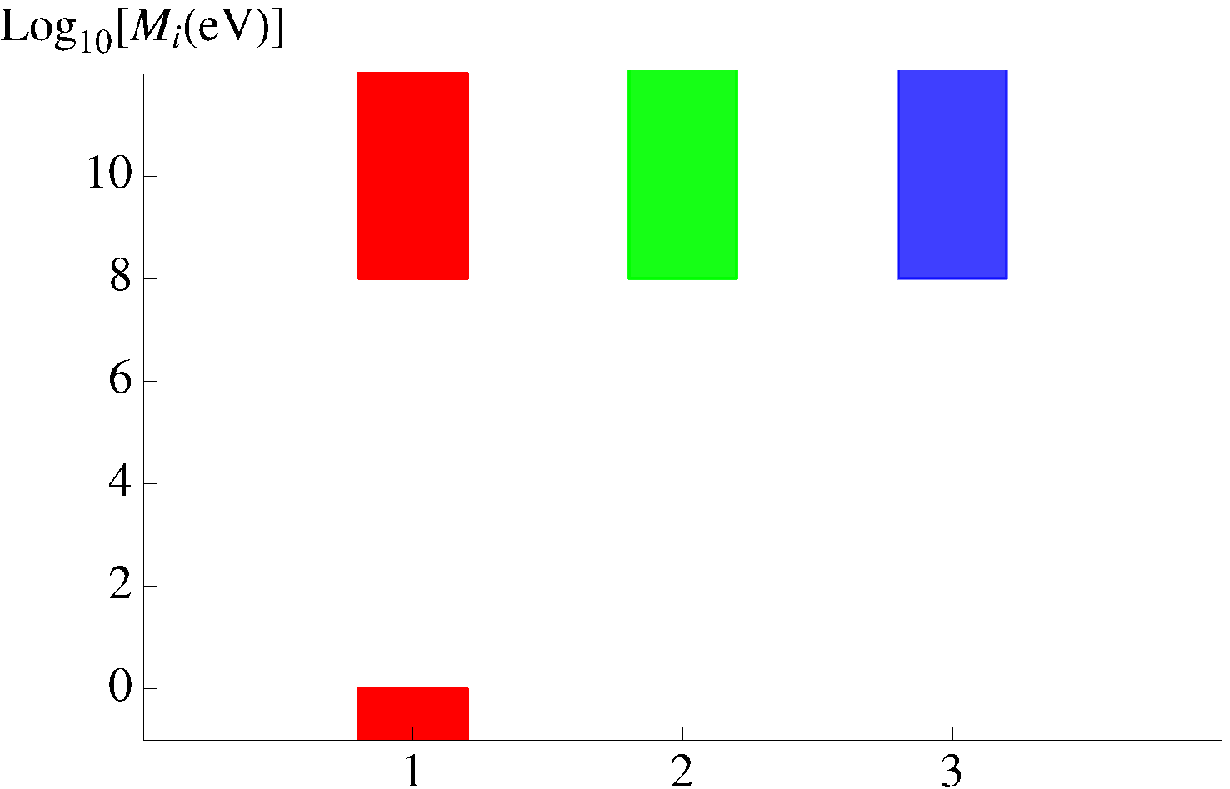}
\includegraphics[width=65mm]{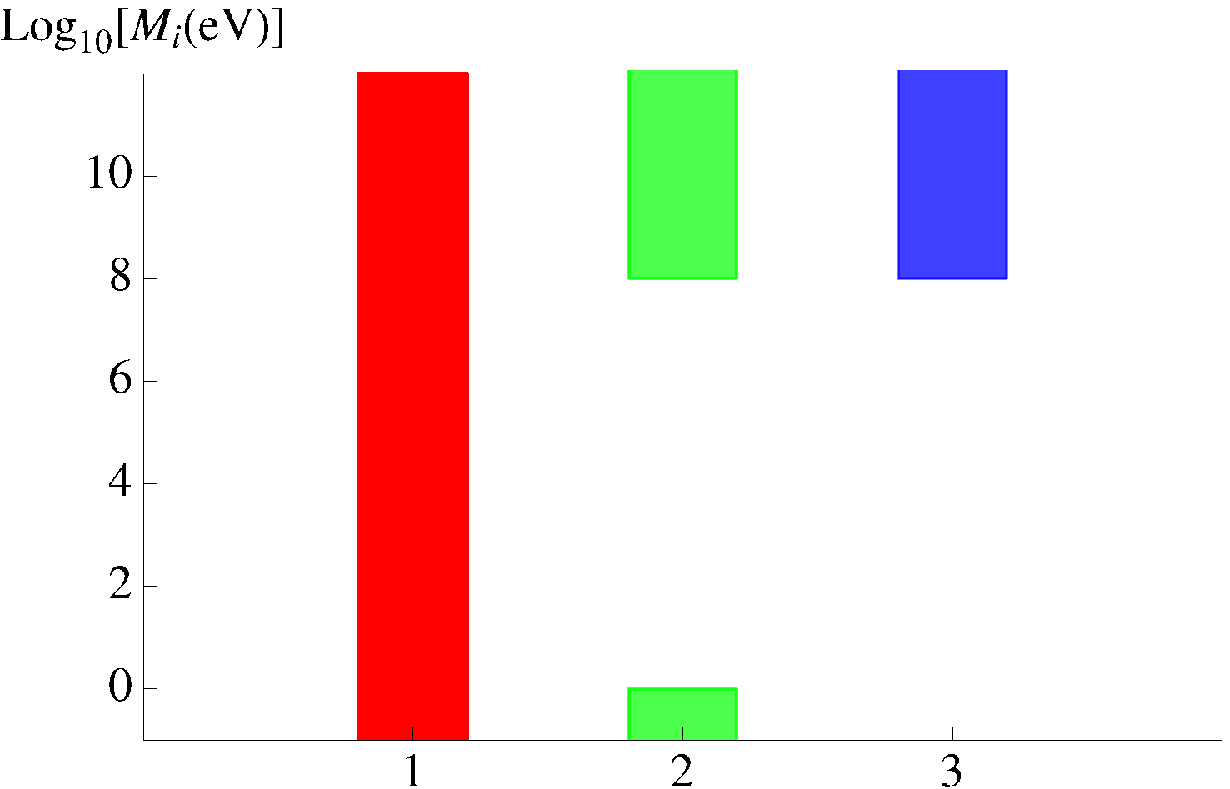}
%}
\end{center}
\caption{
The allowed mass ranges for $n=3$ heavy neutrinos $N_I$ depend upon whether the lightest active neutrino is heavier (left panel) or lighter (right panel)  than $3.25\times 10^{-3}$ eV \cite{Hernandez:2014fha}.
The difference between the two cases comes from a combination of neutrino oscillation data and early universe constraints. 
If the lightest active neutrino is relatively heavy, then all three $N_I$ need to have sizable mixings with active neutrinos to generate the three neutrino masses $m_i$ ($n$ sterile neutrinos that mix with the active neutrinos can generate $n$ active neutrino masses via the seesaw mechanism). This means that all of them are produced in significant amounts in the early universe. In this case there exists a mass range $1 {\rm eV} \lesssim M_I \lesssim 100$ MeV that is excluded for all $N_I$ by cosmological considerations, in particular big bang nucleosynthesis and constraints on the number of effective neutrino species  $N_{\rm eff}$.  
If the lightest active neutrino is massless or rather light, then one heavy neutrino (here chosen to be $N_1$) can have rather tiny mixings $U_{\alpha 1}^2$. This allows to circumvent the cosmological bounds, and $N_1$ can essentially have any mass.  In both cases the upper end of the plot is not an upper bound on the mass; indeed there exists no known upper bound for the $M_I$.
Plot taken from Ref.~\cite{Hernandez:2014fha}.
\label{allowedrange} 
}
\end{figure}

The (type-I) seesaw model is defined by adding $n$ neutral fermions $\nu_R$ with RH chirality to the SM. These are referred to as RH neutrinos because they can couple to the SM neutrinos $\nu_L$ in the same way as the RH and LH part of the charged leptons are coupled.
The Lagrangian reads  
\begin{eqnarray}
	\label{L}
	\mathcal{L} &=&\mathcal{L}_{SM}+ 
	i \overline{\nu_{R}}\slashed{\partial}\nu_{R}-
	\overline{l_{L}}F\nu_{R}\tilde{H} -
	\tilde{H}^{\dagger}\overline{\nu_{R}}F^{\dagger}l_L 
	-{\rm \frac{1}{2}}(\overline{\nu_R^c}M_{M}\nu_{R} 
	+\overline{\nu_{R}}M_{M}^{\dagger}\nu^c_{R}), 
	\end{eqnarray}
where flavour and isospin indices are suppressed.
$\mathcal{L}_{SM}$ is the SM Lagrangian, 
$l_{L}=(\nu_{L},e_{L})^{T}$ are the LH lepton doublets and $H$ is the Higgs doublet with $\tilde{H}=\epsilon H^*$, were $\epsilon$ is the antisymmetric $SU(2)$ tensor. 
$M_{M}$ is a Majorana mass term for $\nu_{R}$ with $\nu_R^c=C\overline{\nu_R}^T$,\footnote{The charge conjugation matrix is $C=i\gamma_2\gamma_0$ in the Weyl basis.} and $F$ is a matrix of Yukawa couplings. 
We work in a flavour basis with $M_M={\rm diag}(M_1,M_2,M_3)$.
For $M_I> 1$ eV there are two distinct sets of mass eigenstates, which we represent by flavour vectors of Majorana spinors $\upnu$ and $N$. The elements $\upnu_i$ of the vector 
\begin{equation}
\upnu=V_\nu^{\dagger}\nu_L-U_\nu^{\dagger}\theta\nu_{R}^c +{\rm c.c.}
\end{equation}
are mostly superpositions of the ``active'' SU(2) doublet states $\nu_L$ and have light masses $\sim -F^2 \times v^2/M_I\ll M_I$. Here c.c. stands for the $c$-conjugation defined above.
The elements $N_I$ of 
\begin{equation}
N=V_N^\dagger\nu_R+\Theta^{T}\nu_{L}^{c} +{\rm c.c.}
\end{equation}
are mostly superpositions of the ``sterile'' singlet states $\nu_R$ with masses of the order of $M_I$. 
At energies below the electroweak scale, these \emph{heavy neutral leptons} interact with the SM via their mixing with active neutrinos, which is characterised by the matrix elements $\Theta_{\alpha I}\ll 1$; if kinematically allowed, they participate in all processes in the same way as SM neutrinos, but with cross sections suppressed by 
\begin{equation}
U_{\alpha I}^2\equiv|\Theta_{\alpha I}|^2\ll 1.
\end{equation}
$V_\nu$ is the usual neutrino mixing matrix $V_\nu\equiv (\mathbbm{1}-\frac{1}{2}\theta\theta^{\dagger})U_\nu$
with  
$\theta\equiv m_D M_M^{-1}$,
where $m_D\equiv Fv$. % and $v$ and $v=174$ GeV.
$U_\nu$ is its unitary part, $V_N$ and $U_N$ are their equivalents in the sterile sector and $\theta\equiv\Theta U_N^T$.
The unitary matrices $U_\nu$ and $U_N$ diagonalise the mass matrices 
\begin{eqnarray}
m_\nu\simeq-v^2FM_M^{-1}F^T=-\theta M_M \theta^T \label{activemass} \ , \
M_N\simeq M_M + \frac{1}{2}\big(\theta^{\dagger} \theta M_M + M_M^T \theta^T \theta^{*}\big), 
\end{eqnarray}
respectively. The eigenvalues of $M_M$ and $M_N$ coincide in very good approximation, and the terms of order $\mathcal{O}[\theta^2]$ are only relevant if two of the $M_I$ are quasi-degenerate.
If (\ref{activemass}) is the only source of neutrino masses, then the $n$ must at least equal the number of non-zero $m_i$, i.e.\ $n\geq 2$ if the lightest neutrino is massless and $n\geq 3$ if it is massive.
Very little is known about the magnitude of the $M_I$.  
If the $N_I$-interactions are to be described by perturbative quantum field theory, then the $M_I$ should be at least 1-2 orders of magnitude below the Planck mass. This can be estimated by inserting the observed neutrino mass differences into (\ref{activemass}).
On the lower end they can have eV (or even sub-eV) masses \cite{deGouvea:2005er}. For $n\geq3$ any value in between is experimentally allowed \cite{Hernandez:2014fha}, see Figs.~\ref{seesawfig} and \ref{allowedrange}. 
In the following I summarise the most popular seesaw scale choices and their phenomenological implications.

\subsection{The GUT-seesaw}
\begin{figure}
\begin{center}
%\vspace{-2cm}
%\uncover<2->{
\includegraphics[width=100mm]{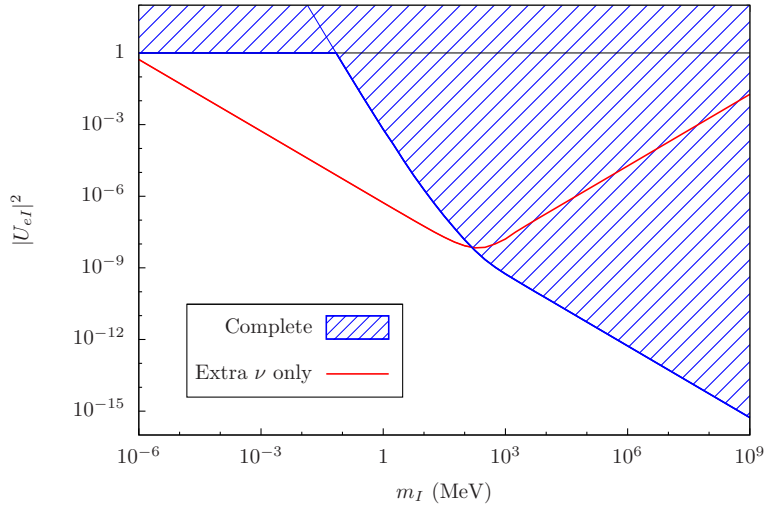}
%}
\end{center}
\caption{If the $N_I$ have a degenerate mass spectrum, then constraints from neutrinoless double $\beta$-decay allow to exclude large parts of the plane spanned by their common mass $m_I$ and $U_{e I}^2$. 
The blue region is excluded if one consistently takes into account the contributions form active and sterile neutrino exchange, the red line was determined under the (incorrect) assumption that only the sterile neutrinos mediate the decay. Plot taken from \cite{Blennow:2010th}.
For a non-degenerate spectrum these bounds are much weaker, see e.g. Fig.~\ref{DlIH}.
\label{0nubbsimple} 
}
\end{figure}

In the probably most discussed version of the seesaw mechanism the $M_I$ are far above the electroweak scale. This choice is primarily theoretically motivated by aesthetic arguments: For "natural" entries (i.e. of order unity) of the $F$, neutrino masses near the upper limit on $\sum_i m_i< 0.23$ imposed by Planck \cite{Ade:2013zuv} imply values of $M_I\sim 10^{14}-10^{15}$ GeV, slightly below the  suspected scale of grand unification. 
Hence, this scenario can easily be embedded in grand unifying theories. 
Heavy $M_I$ are also well-motivated from cosmology.
Leptogenesis \cite{Fukugita:1986hr} is one of the most popular explanations for the observed BAU (2).
In the most studied version of leptogenesis, the BAU is generated in the CP-violating decay of $N_I$, see e.g. \cite{Buchmuller:2005eh} for a review. For a non-degenerate mass spectrum and without any degrees of freedom in addition to (\ref{L}), this mechanism only works for $M_1\gtrsim4\times10^{8}$ GeV \cite{Davidson:2002qv}.
Flavour effects \cite{Barbieri:1999ma} 
can reduce this lower bound by $1-2$ orders of magnitude \cite{Antusch:2009gn}, which is still far out of experimental reach.\footnote{The consistent description of all quantum and flavour effects remains an active field of research \cite{Cirigliano:2009yt}.}
If the $M_I$ are indeed that large, then the new states $N_I$ cannot be found in any near future experiment (and possibly never). 
The only traces they leave in experiments can be parametrised in terms of higher dimensional operators in an effective Lagrangian obtained after integrating them out \cite{Broncano:2002rw}, including the Weinberg operator (\ref{Weinberg}) with $f=F M_M^{-1}F^T$.
On the positive side, probing these operators allows to constrain some of the parameters in $F$ and $M_M$ by looking for rare processes, such as neutrinoless double $\beta$-decay or $\mu\rightarrow e\gamma$, hence indirectly testing physics at a very high scale. For a degenerate $M_I$-spectrum or $n=1$ these bounds can indeed be quite strong, 
see Fig.~\ref{0nubbsimple}; for $n=3$ they are much weaker \cite{Drewes:2015iva}. 
On the negative side, the seesaw mechanism is not the only way to generate these operators, and without directly finding the new states, it is impossible to definitely distinguish it from other scenarios. While leptogenesis may be falsified indirectly \cite{Deppisch:2013jxa}, the GUT seesaw itself can always escape falsification if the CP-violation is small and the BAU generated from another source. This lack of falsifiability is somewhat unsatisfying, though Nature might not care about this.
Moreover, with couplings $F$ of order unity, the $N_I$ contribute to the hierarchy problem, 
and thermal leptogenesis requires a large reheating temperature, which is at tension with upper limits on the temperature in supersymmetric theories \cite{Pagels:1981ke} ("gravitino problem").   

\subsection{The TeV-seesaw} 
\begin{figure}
  \centering
    \includegraphics[width=12cm]{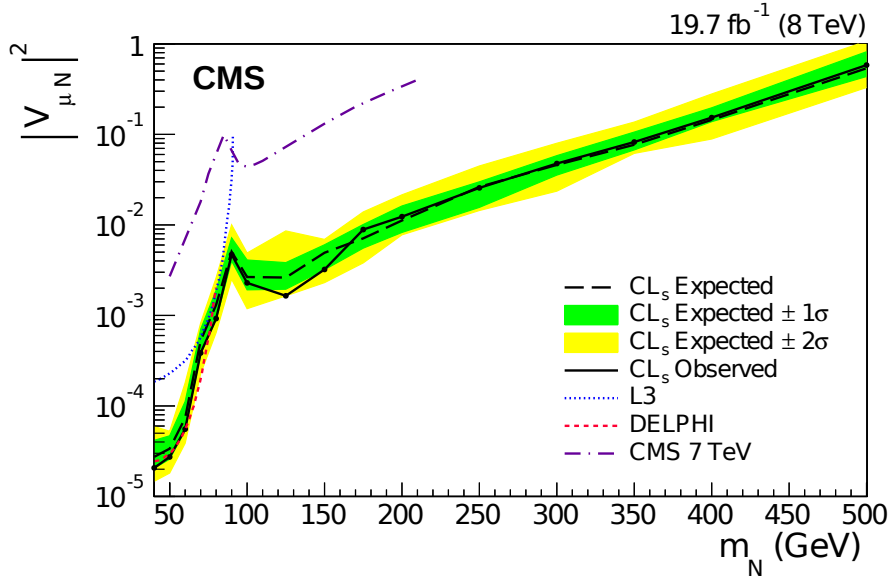}
    \caption{Constraints on $U_{\mu I}^2$ (here referred to as $|V_{\mu N}|^2$) as a function of $M_I$ (here referred to as $m_N$) from the CMS experiment (solid black line). Plot taken from Ref.~\cite{Khachatryan:2015gha}. 
\label{CMS}}
\end{figure}

The highest scale that can be probed directly by collider experiments is the TeV scale, see e.g. \cite{Deppisch:2015qwa} for a recent overview. 
Searches for heavy neutral leptons like $N_I$ have been undertaken at the ATLAS and CMS experiments at the Large Hadron Collider (LHC) \cite{ATLAS:2012ak,Khachatryan:2014dka,Khachatryan:2015gha}, see Fig.~\ref{CMS},  so far without positive result. These searches have been performed for both, the minimal seesaw (\ref{L}) as well as its left-right symmetric extension, see e.g. \cite{Barry:2013xxa}. 
The experimental challenge lies mainly in the fact that the Yukawa interactions $F$ that govern the branching ratios are constrained by the seesaw relation (\ref{activemass}); a relatively low seesaw scale $M_I$ at or below the TeV scale generally requires very small values of the $F_{\alpha I}\sim F_0\equiv(m_i M_I/v^2)^{1/2}$, hence unobservable tiny branching ratios. 
In the minimal seesaw (\ref{L}) a discovery at the LHC is only realistic if the individual $F_{\alpha I}$ are much bigger than $F_0$, and the smallness of the $m_i$ is achieved due to a cancellation in the matrix valued equation (\ref{activemass}) \cite{Kersten:2007vk,Ibarra:2011xn,Das:2012ze}. This is realised in models with approximate lepton number conservation, and chances are generally better in extensions of (\ref{L}) in which the $N_I$ have additional interactions, see \cite{Mohapatra:1986bd,Chikashige:1980ui}.

If the $M_I$ are slightly lower, below the masses of the W and Z-boson, then $N_I$ can be produced in the decay of these gauge bosons.\footnote{Production is also possible in Higgs decays \cite{Cely:2012bz}.} This allows to impose much stronger constraints \cite{Abreu:1996pa}, see Figs.~\ref{UePlot}-\ref{UtauPlot} and \cite{Drewes:2015iva} for a summary.
Possible future direct searches  have e.g. been studied in \cite{Blondel:2014bra}.
In addition to direct searches, neutrino oscillation experiments and the bounds on low energy lepton flavour violation (LFV)\footnote{The most sensitive low energy processes are $\mu\rightarrow e\gamma$ decays and muon to electron conversion in nuclei \cite{Alonso:2012ji}. Experimentally these are different from LFV decays of heavier particles in colliders, which are amongst the typical signatures for $N_I$ in ``direct searches''.} and lepton number violation \cite{LopezPavon:2012zg,Atre:2009rg} mentioned in the previous paragraph as well as electroweak precision data impose indirect constraints on the $N_I$-properties, see e.g. \cite{Atre:2009rg,Akhmedov:2013hec}.

Heavy neutrinos $N_I$ with $M_I$ at the electroweak or TeV scale are also interesting cosmologically because they can generate the BAU via leptogenesis either during their decay \cite{Pilaftsis:2003gt}  or thermal production ("baryogenesis from neutrino oscillations") \cite{Akhmedov:1998qx}. 
For $n=2$ flavours of sterile neutrinos the observed BAU can only be explained in the minimal model (\ref{L}) if it is enhanced by a mass degeneracy \cite{Pilaftsis:2003gt}, for $n=3$ or more flavours no degeneracy is required \cite{Garbrecht:2014bfa}.
This is further relaxed in models with additional degrees of freedom \cite{Kang:2014mea,Bezrukov:2012as}.
Though leading order estimates exist, the quantitative description of these low scale leptogenesis scenarios is highly non-trivial due to the complicated interplay of quantum, thermodynamic and flavour effects and remains an active field of research \cite{Garny:2011hg}.

\begin{figure}
\begin{center}
\includegraphics[width=12cm]{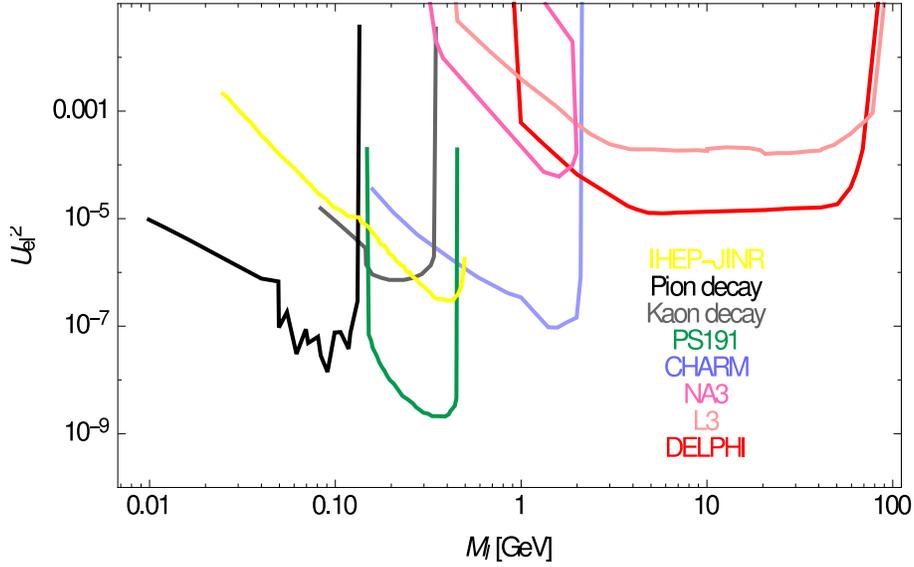}
\caption{\label{UePlot}
Constraints on the mixing $U_{e I}^2$ from the experiments 
DELPHI \cite{Abreu:1996pa}, L3 \cite{Adriani:1992pq}, 
PIENU \cite{PIENU:2011aa},
TRIUMF/TINA \cite{Britton:1992xv},
PS191 \cite{Bernardi:1987ek},
CHARM \cite{Bergsma:1985is},
NA3  \cite{Badier:1985wg}, %as estimated in Pascoli
IHEP-JINR \cite{Baranov:1992vq}
and kaon decays \cite{Yamazaki:1984sj}. %and other list.
The plot is similar to Ref.~\cite{Drewes:2015iva}, some comments on the interpretation can be found in that article and references therein.\label{UePlot}
}
\end{center}
\end{figure}
\begin{figure}
\begin{center}
\includegraphics[width=12cm]{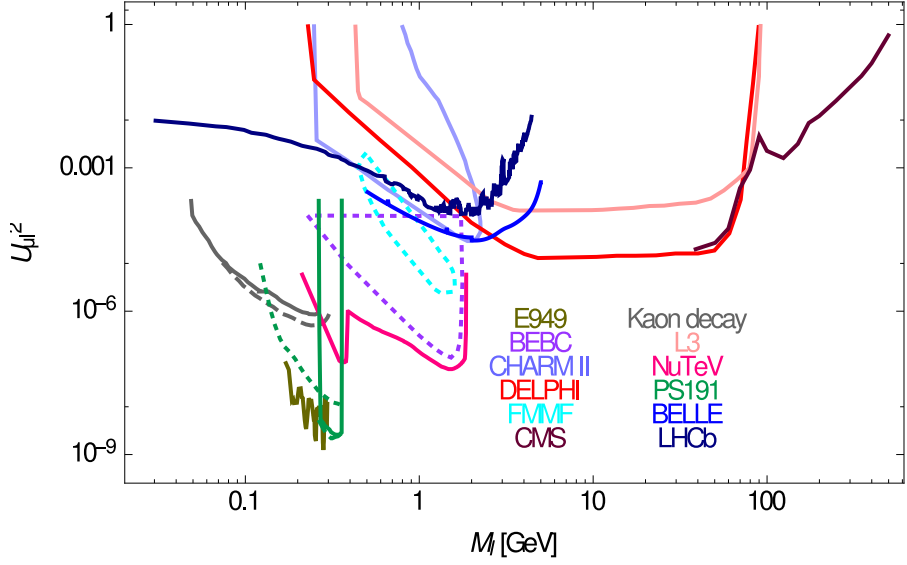}
\caption{\label{UmuPlot}
Constraints on the mixing $U_{\mu I}^2$ from the experiments CMS \cite{Khachatryan:2015gha}, DELPHI \cite{Abreu:1996pa}, L3 \cite{Adriani:1992pq}, LHCb \cite{Aaij:2014aba}, BELLE \cite{Liventsev:2013zz}, 
BEBC \cite{CooperSarkar:1985nh}, %as in Pascoli
FMMF \cite{Gallas:1994xp}, %as in Pascoli
E949 \cite{Artamonov:2014urb},
PIENU \cite{PIENU:2011aa}, 
TRIUMF/TINA \cite{Britton:1992xv},
PS191 \cite{Bernardi:1987ek},  
CHARMII \cite{Vilain:1994vg},
NuTeV \cite{Vaitaitis:1999wq}, 
NA3 \cite{Badier:1985wg}
and kaon decays in \cite{Yamazaki:1984sj,Hayano:1982wu}.
The plot is similar to Ref.~\cite{Drewes:2015iva}, some comments on the interpretation can be found in that article and references therein.
}
\end{center}
\end{figure}
\begin{figure}
\begin{center}
\includegraphics[width=12cm]{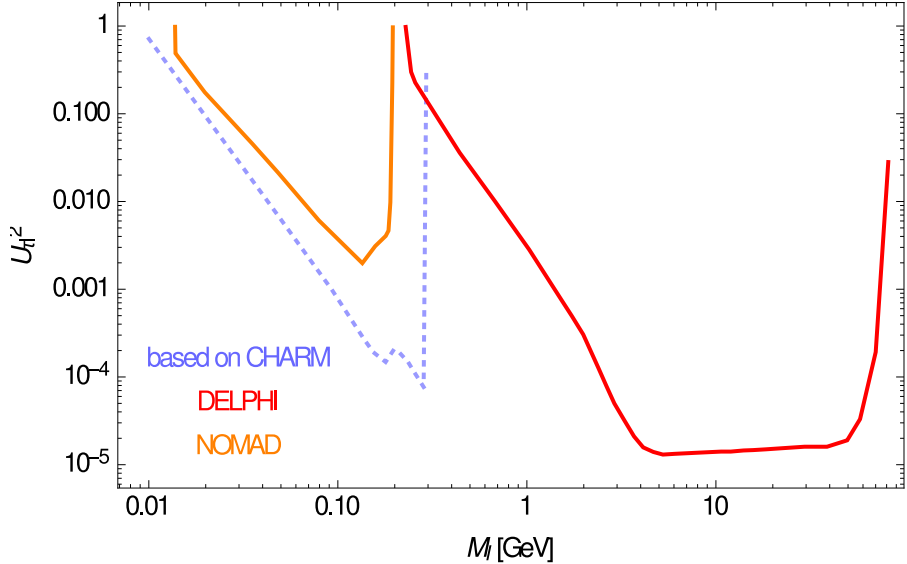}
\caption{\label{UtauPlot}
Bounds on the mixing $U_{\tau I}^2$ based on the interpretation of CHARM data in \cite{Orloff:2002de}, NOMAD \cite{Astier:2001ck} and DELPHI \cite{Abreu:1996pa}. Plot taken from Ref.~\cite{Drewes:2015iva}.
The plot is similar to Ref.~\cite{Drewes:2015iva}, some comments on the interpretation can be found in that article and references therein. The bounds may be considerably improved when studying $\tau$-decays in future b-factories \cite{Kobach:2014hea}.\label{UtauPlot}
}
\end{center}
\end{figure}

\begin{landscape}
\begin{figure}
\begin{center}
\begin{tabular}{c c}
\includegraphics[width=10cm]{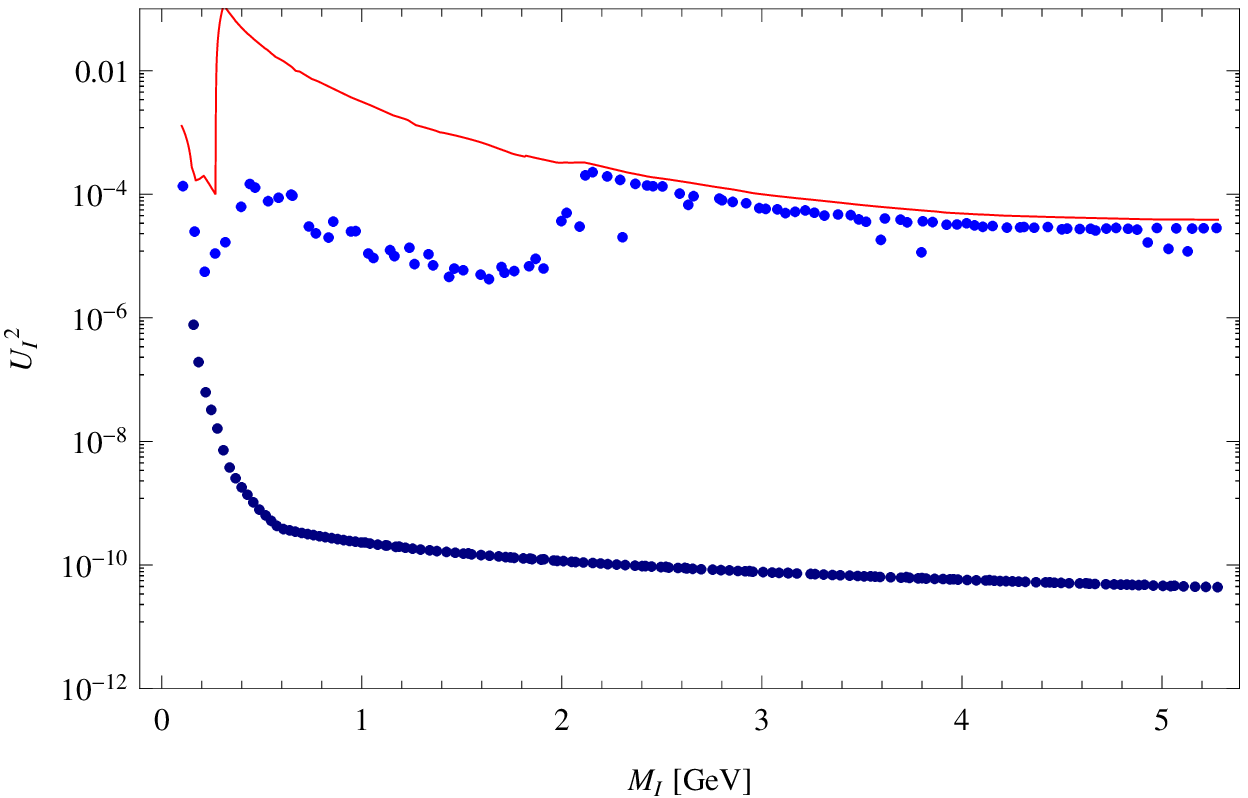}
&
\includegraphics[width=10cm]{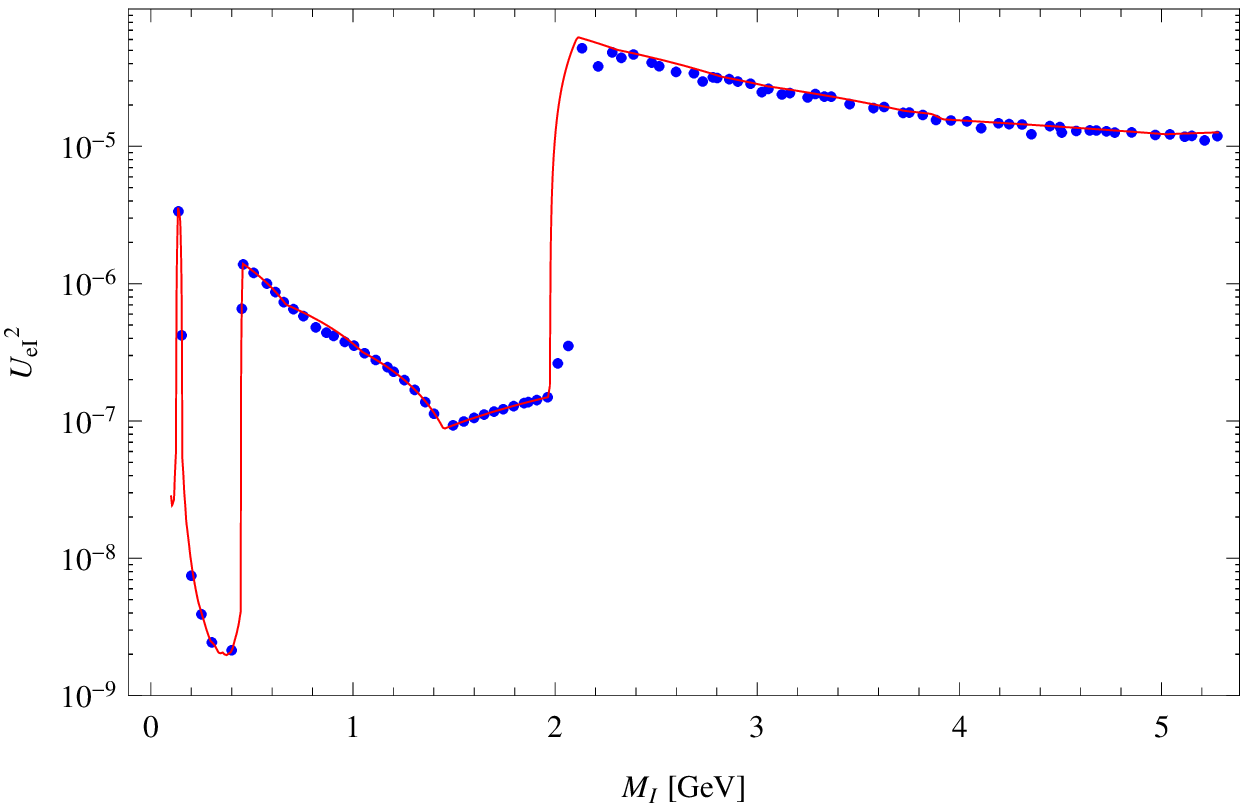}
\\
\includegraphics[width=10cm]{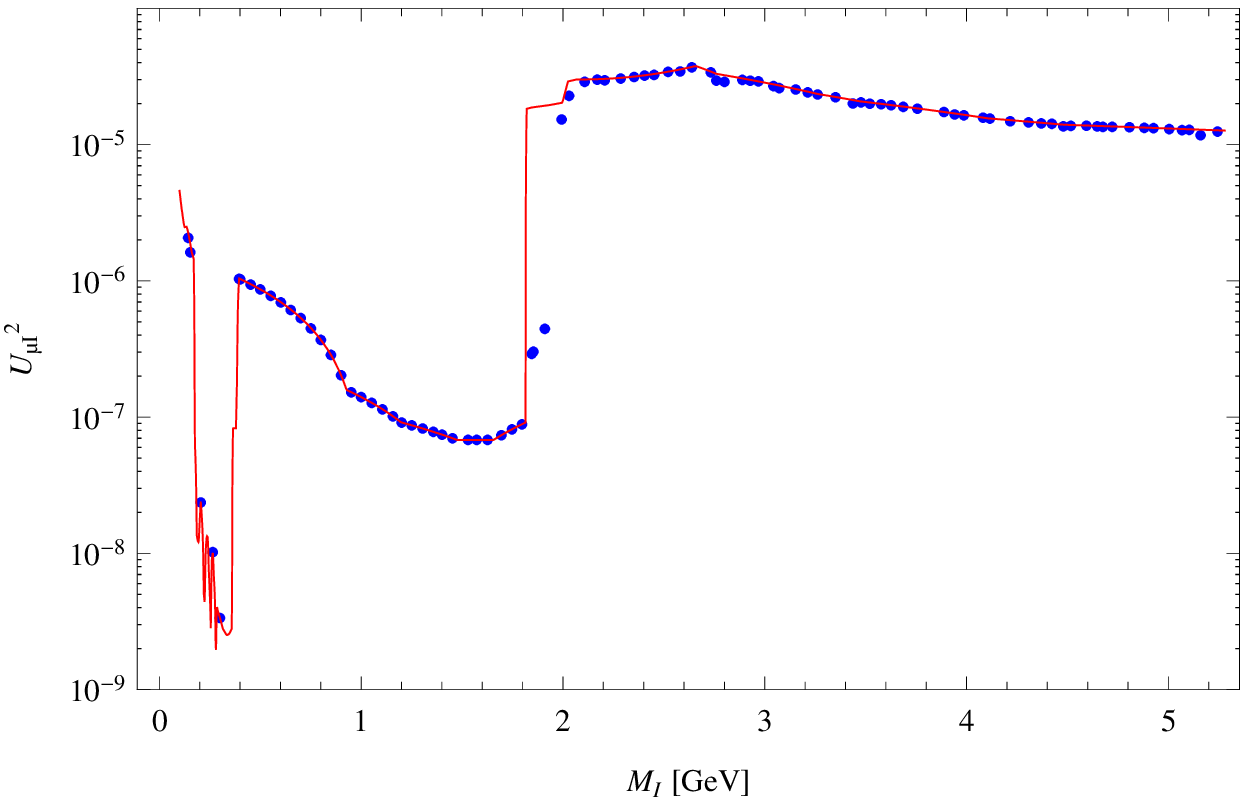}
&
\includegraphics[width=10cm]{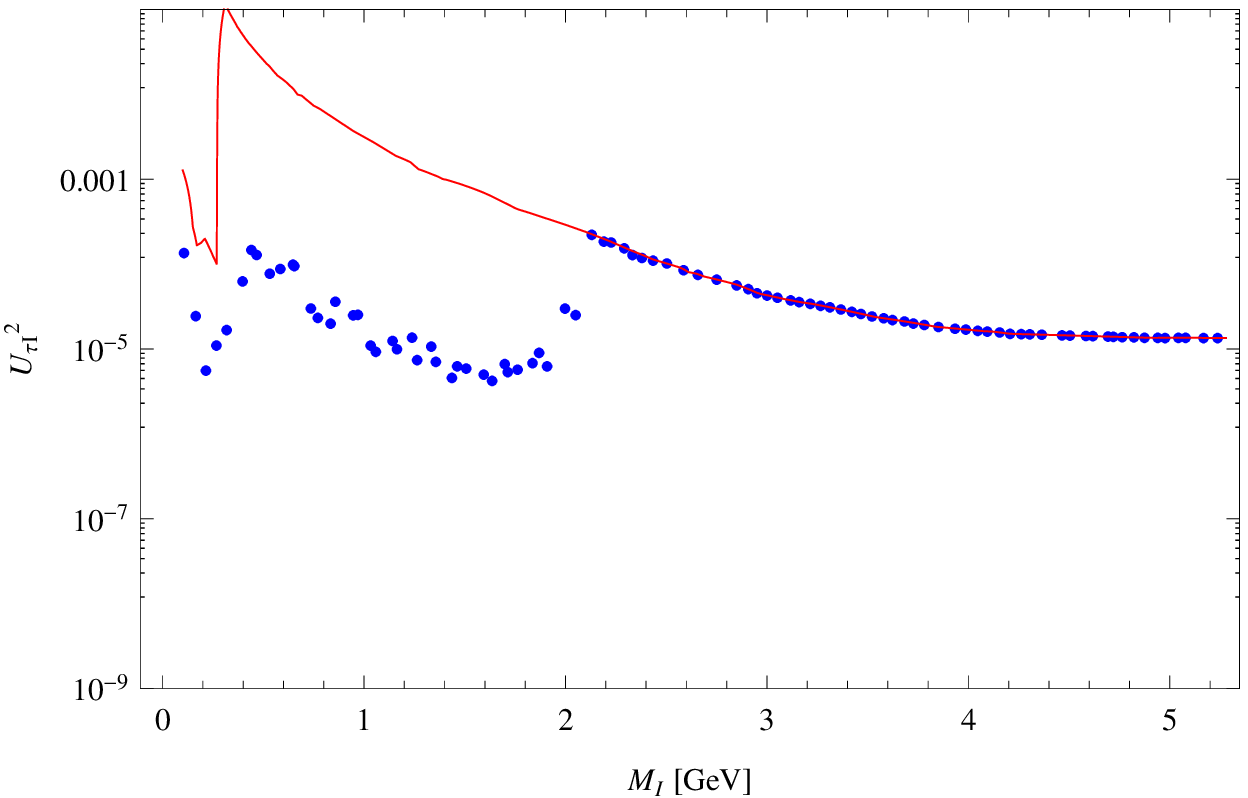}
\end{tabular}
\end{center}
\caption{
Constraints on mass and mixing for a seesaw with $n=3$ heavy neutrinos 
with inverted $m_i$-hierarchy with $m_3=0.23$ eV.
The light blue dots in the upper left panel show the largest  value of $U_I^2=\sum_\alpha U_{\alpha I}^2$ for a given $m_\pi<M_I<m_B$ found to be consistent with collider searches, neutrino oscillation data, $\mu\rightarrow e \gamma$ searches, neutrinoless double $\beta$-decay and big bang nucleosynthesis in Ref.~\cite{Drewes:2015iva}. 
The masses of the other two heavy neutrinos were allowed to vary between the pion mass $m_\pi$ and the mass $m_W$ of the W-boson. 
The red line is the ``naive'' sum of the upper bounds on $U_{e I}^2$, $U_{\mu I}^2$ and $U_{\tau I}^2$ from direct searches obtained from the direct search experiments shown in Figs.~\ref{UePlot}-\ref{UtauPlot}. 
The other panels show the same for the individual $U_{\alpha I}^2$.
The dark blue dots show the lower limit on $U_I^2$ from neutrino oscillation data and big bang nucleosynthesis. The lower limit is considerably weaker if the lightest neutrino is massless \cite{Drewes:2015iva}. 
\label{DlIH}
}
\end{figure}
\end{landscape}

\subsection{The GeV-seesaw}
For $M_I$ below the mass of the B-mesons the existing constraints and the perspectives for future searches improve significantly (and even more below the D-meson mass), see \cite{Atre:2009rg,Drewes:2015iva} for a comprehensive overview. 
On one hand the $N_I$ can be produced efficiently in meson decays, allowing to impose upper bounds on the individual $U_{\alpha I}^2$ shown in Figs.~\ref{UePlot}-\ref{UtauPlot}, see Refs.~\cite{Gorbunov:2007ak,Atre:2009rg,Boyarsky:2009ix,Drewes:2013gca,Canetti:2014dka,Drewes:2015iva} and references therein for more details.
On the other hand neutrino oscillation data and the seesaw relation (\ref{activemass}) impose stronger bounds on the sum 
\begin{equation}
U^2_I\equiv\sum_\alpha U_{\alpha I}^2. 
\end{equation}
Finally, the requirement to decay before BBN \cite{Hernandez:2014fha} imposes a strict lower bound  on $U^2_I$ as a function of $M_I$ \cite{Ruchayskiy:2012si,Drewes:2015iva}.
There are constraints from lepton flavour violation \cite{Canetti:2013qna,Drewes:2015iva} neutrinoless double $\beta$-decay \cite{Bezrukov:2005mx,LopezPavon:2012zg,Drewes:2015iva} and lepton universality in meson decays \cite{Abada:2012mc,Drewes:2015iva}. For $N_I$ heavier than D-mesons, they can be more constraining than direct search bounds \cite{Drewes:2015iva}.

\begin{figure}[h!]
%\centering
%\psfrag{M}{$M_2$ [GeV]}
%\psfrag{U}{$U_\mu^2$}
\centering
\includegraphics[width=12cm]{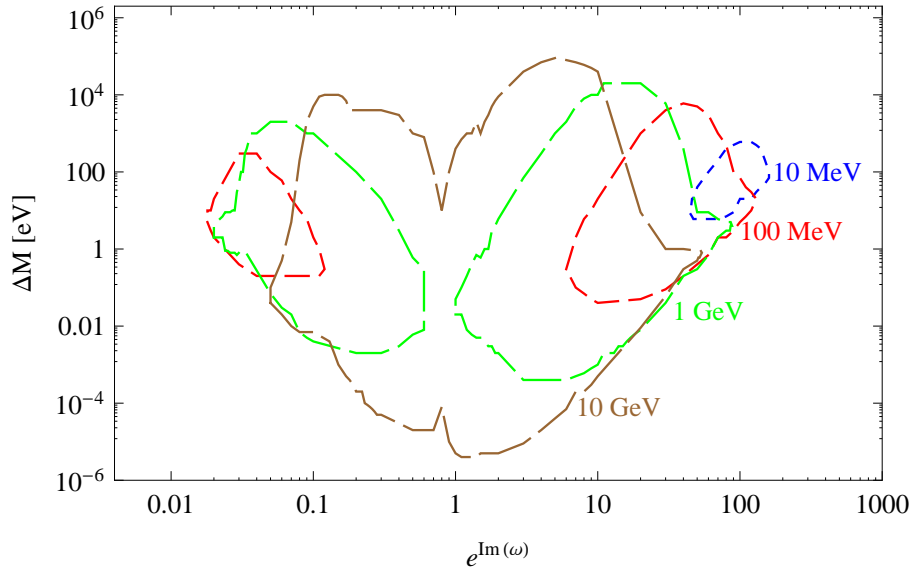}
\caption{It is possible to generate the BAU with $n=2$ heavy neutrinos with masses below the electroweak scale if these have degenerate masses $M\equiv (M_1+M_2)/2\gg \Delta M\equiv |M_1-M_2|/2$. For $n=3$, no such degeneracy is needed, see Fig.~\ref{bounds}.
This low scale leptogenesis scenario relies on CP-violating neutrino oscillations \cite{Akhmedov:1998qx}.
However, in general the CP-violation does not come from the phases in the neutrino mixing matrix $U_\nu$, but from another CP-violating parameter in the sterile sector (here called ${\rm Im }\omega$). 
This plot shows the regions in the $\Delta M$-$e^{{\rm Im }\omega}$ plane where the BAU can be explained for $n=2$ and normal hierarchy.
Interestingly, for $M>1$ GeV it is possible to generate the BAU from the CP-violation in $U_\nu$ alone (${\rm Im }\omega=0$, along the vertical line in the middle of the plot). In principle the CP-violation due to ${\rm Im }\omega$ can also be measured in meson decays \cite{Cvetic:2014nla}. Plot taken from Ref.~\cite{Canetti:2012vf}.
\label{CPviolation}
}
\end{figure}

\begin{figure}[h!]
%\centering
%\psfrag{M}{$M_2$ [GeV]}
%\psfrag{U}{$U_\mu^2$}
\centering
\includegraphics[width=12cm]{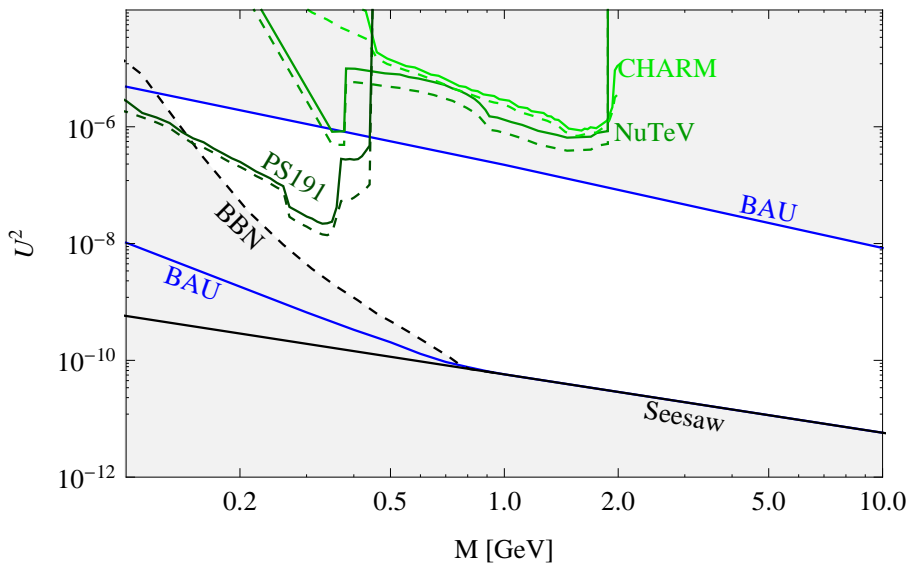}
\caption{Constraints on the $n=2$ scenario with normal hierarchy. For successful leptogenesis, the two $N_I$ must have degenerate masses $M_1\simeq M_2 \simeq M\equiv (M_1+M_2)/2$, see figure \ref{CPviolation}.
The BAU can be explained between the two blue lines.
Points below the ``seesaw'' line are excluded by neutrino oscillation data, points below the ``BBN'' line are excluded by the requirement that the $N_I$ lifetime must be short enough that they do not spoil the agreement of big bang nucleosynthesis predictions with observed light element abundances.
The green lines indicate upper bounds from searches prior to 2012.
The minimal $n=2$ scenario predicts that the lightest active neutrino is massless. Plot taken from Ref.~\cite{Canetti:2012vf}.
\label{nuMSM}
}
\end{figure}

\begin{figure}[h!]
%\centering
%\psfrag{M}{$M_2$ [GeV]}
%\psfrag{U}{$U_\mu^2$}
\centering
\includegraphics[width=12cm]{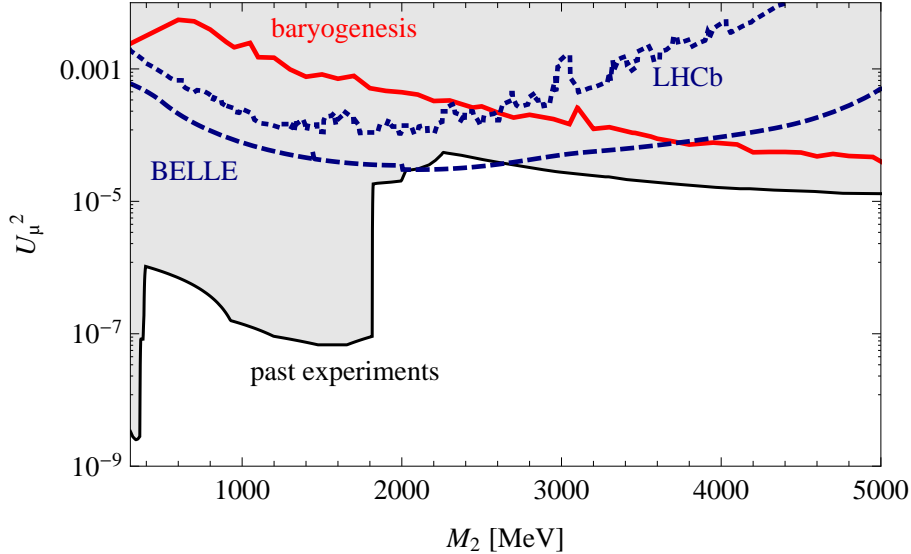}
\caption{The red line shows the maximal mixing $|\Theta_{\mu 2}|^2$ found to be consistent with baryogenesis in Ref.~\cite{Canetti:2014dka} for $n=3$, normal hierarchy and fixed $M_1=1$ GeV and $M_3=3$ GeV. Below the line there exist parameter choices for which the observed BAU can be generated.
The grey area represents bounds from the past experiments PS191 \cite{Bernardi:1987ek}, NuTeV \cite{Vaitaitis:1999wq} (both re-analysed in \cite{Ruchayskiy:2011aa}), NA3 \cite{Badier:1985wg}, CHARMII \cite{Vilain:1994vg} and DELPHI \cite{Abreu:1996pa}. 
The blue lines indicate the current bounds from LHCb \cite{Aaij:2014aba} (dotted) and BELLE \cite{Liventsev:2013zz} (dashed), which will improve in the future. 
\label{bounds}
}
\end{figure}

For $M_I$ in the GeV range, the BAU can be explained via leptogenesis during the thermal production of the $N_I$  \cite{Akhmedov:1998qx,Asaka:2005pn,Canetti:2010aw,Canetti:2012vf,Shuve:2014zua}, see Figs.~\ref{CPviolation}-\ref{bounds}. Similarly to the TeV scale, this requires a mass degeneracy for $n=2$ \cite{Canetti:2012vf,Shuve:2014zua}, see Fig.~\ref{CPviolation}.
No such degeneracy is needed for $n=3$ \cite{Drewes:2012ma}, see Fig.~\ref{bounds}. The leptogenesis parameter space will be further explored in the near future. For $M_I$ below the D-meson mass this is e.g.\ done by the NA62 experiment, for heavier masses LHCb and BELLE II will improve the bounds \cite{Canetti:2014dka}. 
The ideal tool to search for $N_I$ in the GeV range would be the proposed SHiP experiment \cite{Bonivento:2013jag}.
In a small fraction of the parameter space the CP-violation responsible for the BAU comes from the phases in $U_\nu$ that may be measured in neutrino oscillation experiments \cite{Canetti:2012vf}, see Fig.~\ref{bounds}, but in general it lies in the sterile sector and can only be measured in $N_I$ decays if their mass spectrum is degenerate \cite{Cvetic:2014nla}.

\subsection{The keV-seesaw}
\begin{figure}[h!]
%\centering
%\psfrag{M}{$M_2$ [GeV]}
%\psfrag{U}{$U_\mu^2$}
\centering
\includegraphics[width=10cm]{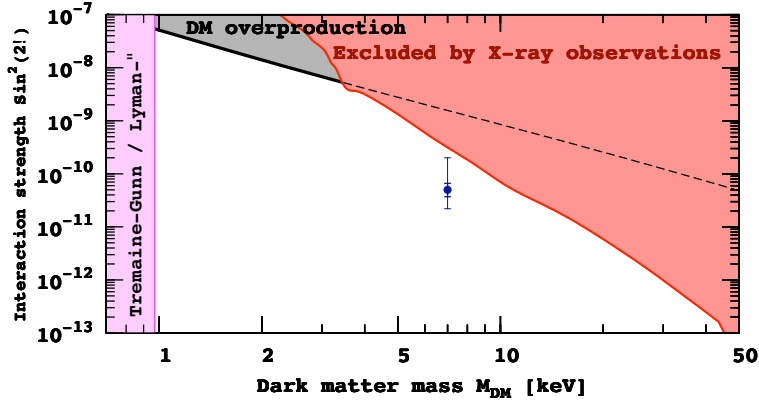}
\caption{Some constraints on the mass and mixing for sterile neutrino DM.
The red region is excluded by the X-ray bounds prior to 2014, see e.g. Fig.~12 in Ref.~\cite{Drewes:2013gca} and references therein for details. 
The dot marks the interpretation of the unexplained 3.5 keV emission in terms of sterile neutrino DM \cite{Bulbul:2014sua}. 
The region on the left is excluded by phase space considerations \cite{Gorbunov:2008ka}.
The solid black \emph{production curve} marks the combinations of mass and mixing for which the the observed 
DM density is explained if sterile neutrinos are only produced thermally via their mixing \cite{Dodelson:1993je}.
The thermal production can be enhanced by lepton asymmetries in the primordial plasma \cite{Shi:1998km}, which can be much bigger than the BAU (see e.g. Ref.~\cite{Canetti:2012zc} for a short summary of the known constraints on the lepton asymmetries and collection of references). This makes it possible to explain the observed DM with mixing angles below the black line. Non-thermal production mechanisms also generate the observed DM density for mixings below this line, see e.g. \cite{Shaposhnikov:2006xi,Bezrukov:2008ut,Petraki:2007gq,Gorbunov:2010bn}. 
We do not display a lower bound on the DM mass from structure formation because it strongly depends on the production mechanism.
For thermal production it should lie somewhere between $1$ keV and $10$ keV \cite{Boyarsky:2008xj}, but this is a matter of ongoing discussion. 
All these bounds assume that 100\% of the DM is made of sterile neutrinos.
Plot similar to  Boyarsky et. al. in Ref.~\cite{Bulbul:2014sua}.
\label{DMfig}
}
\end{figure}
\begin{figure}[h!]
%\centering
%\psfrag{M}{$M_2$ [GeV]}
%\psfrag{U}{$U_\mu^2$}
\centering
\includegraphics[width=9cm]{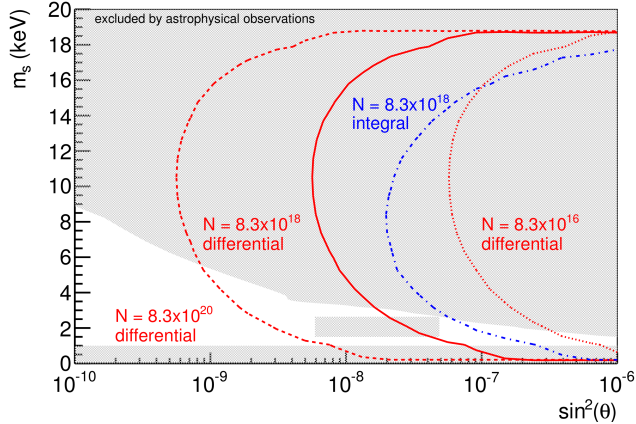}
\caption{The estimated sensitivity of the KATRIN experiment to keV mass sterile neutrinos after 3 years of measurement, as estimated in Ref.~\cite{Mertens:2014nha}. Here the sterile neutrino's mass $M_I$ is denoted by $m_s$, and N refers to the effective number of tritium atoms in the source.
The perspectives have also been studied in left-right symmetric theories \cite{Barry:2014ika}.\label{KatrinFig}
}
\end{figure}
\begin{figure}
  \centering
    \includegraphics[width=9cm]{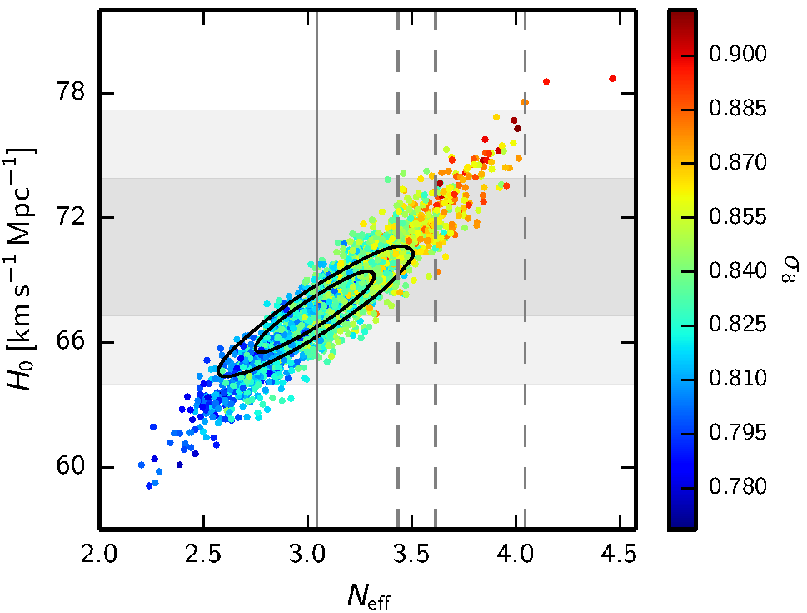}
    \caption{Constraints on the Hubble constant $H_0$ and the number of relativistic degrees of freedom $N_{\rm eff}=3.046 + \Delta N_{\rm eff}$ \cite{Mangano:2001iu} in the primordial plasma at CMB decoupling, as presented in Ref.~\cite{Planck:2015xua}. 
A light sterile neutrino in thermal equilibrium would imply $\Delta N_{\rm eff} =1$. Such a high value is over $3\sigma$ away from best fit marked by the dark contours, which is extracted from the combination of Planck data and baryonic acoustic oscillations (BAO). 
This clearly disfavours light sterile neutrinos with masses and mixings suggested by fits to the oscillation anomalies. The tension can be relaxed if one allows $H_0$ to depart from its best fit value $H_0= 67.8 \pm 0.)$ km/s/Mpc, which indeed seems to be favoured by some local measurements of $H_0$. This, however, comes at the cost of moving the normalisation of the linear power spectrum away from its best fit value $\sigma_8 =0.829 \pm 0.015$ (colour coding).
A light sterile neutrino that is not in thermal equilibrium would give a contribution $\Delta N_{\rm eff} <1$, which is still allowed by CMB constraints (and also BBN constraints, see e.g. \cite{Barger:2003zg,Canetti:2012zc} and references therein).
Different interpretations of this situation can e.g. be found in Refs.~\cite{Leistedt:2014sia}.
\label{planck}
}
\end{figure}
Sterile neutrinos $N_I$ are massive, feebly interacting and can be very long lived.
This makes them obvious DM candidates \cite{Dodelson:1993je,Shi:1998km,Abazajian:2001nj}.
Their properties are constrained by astrophysical, cosmological and laboratory data. Most importantly, the radiative decay $N\rightarrow \nu \gamma$ would lead to an observable photon emission line at energy $M_I/2$ from DM dense regions \cite{Abazajian:2001vt}. Until 2014, the non-observation of an emission line could only be use to impose an upper bound on $U_I^2$ as a function of $M_I$, see Fig.~\ref{DMfig}. These ``established constraints'' are e.g.\ discussed in \cite{Boyarsky:2012rt,Drewes:2013gca} and references therein; 
they imply that the $N_I$ that compose the DM must be so feebly coupled that they cannot contribute significantly to the neutrino masses (\ref{activemass}) or leptogenesis.
In 2014, two groups reported an unexplained emission signal at $\sim 3.5$ 
keV \cite{Bulbul:2014sua} that can be interpreted as evidence for sterile neutrino DM, 
though this interpretation is disputed \cite{Jeltema:2014qfa}. 
Observations with the Astro-H satellite \cite{Takahashi:2012jn} may help to clarify the situation. Since thermal production via mixing is unavoidable \cite{Dodelson:1993je}, an upper bound on $U_I^2$ can also be obtained from the requirement not to produce too much DM. 
The DM mass $M_I$ is bound from below by phase space considerations (essentially Pauli's exclusion principle in DM dense regions) \cite{Gorbunov:2008ka}.
In the laboratory, $U_I^2$ can be constrained by KATRIN and similar experiments \cite{Mertens:2014nha}, see Fig.~\ref{KatrinFig}.
It has also been pointed out that keV mass sterile neutrinos are responsible for pulsar kicks \cite{Kusenko:1998bk} and affect supernova explosions \cite{Hidaka:2006sg}.\footnote{Supernova constraints also exist for heavier neutrinos \cite{Dolgov:2000pj}, but at least in the minimal model (\ref{L}) they are generally weaker than the combination of BBN and laboratory constraints.}

All other constraints depend on the mechanism by which the $N_I$ are produced in the early universe. 
Thermal production via mixing is most efficient at temperatures $T\sim 100$ MeV \cite{Asaka:2006nq} and always leads to a $N_I$ population with a thermal momentum distribution, but a total abundance far below the equilibrium value. X-ray bounds force $M_I$ to be in the keV range, which can be realised in different models \cite{Merle:2013gea}.
This "warm DM" component cannot compose all the DM, as their mean free path during structure formation in the early universe would be in tension with the observed small scale structure in the universe, see e.g. \cite{Boyarsky:2012rt,Drewes:2013gca} and references therein.
However, in the presence of a significant lepton asymmetry, there is also a resonantly produced "cold" component \cite{Shi:1998km} produced due to the MSW effect, which allows to explain all DM in terms of sterile neutrinos in agreement with structure formation bounds \cite{Boyarsky:2008xj}. 
Interestingly, the three phenomena (1)-(3) can be explained simultaneously within the minimal model (\ref{L}) if one of the $N_I$ has a keV mass and acts as DM while the other two have degenerate masses in the GeV range and are responsible for leptogenesis and neutrino masses \cite{Boyarsky:2009ix}. This scenario, first proposed in \cite{Asaka:2005pn}, was shown to be feasible in \cite{Canetti:2012vf}. A cold component can also be produced non thermally, e.g. due to a coupling to an inflaton \cite{Shaposhnikov:2006xi}, the SM Higgs \cite{Bezrukov:2008ut}, other scalars \cite{Petraki:2007gq} or modified gravity \cite{Gorbunov:2010bn}. 
Then the mass can be larger than keV because the production for not rely on the mixing $\theta$, which can in turn be made arbitrarily small to ensure a long lifetime even for a large mass.
The spectrum can also effectively be "cooled down" if entropy is injected into the rest of the primordial plasma due the decay of some heavy particle, which allows to reconcile an initially  warm or hot DM spectrum with structure formation constraints \cite{Bezrukov:2009th}.

\subsection{The (sub)eV-seesaw} 
In principle neutrino oscillation data can be explained via the seesaw mechanism with $M_I$ as low as an eV \cite{deGouvea:2005er}. 
For even lower masses the seesaw hierarchies $M_I\gg m_i$ and $\theta\ll1$ do not hold, and for $M_I\ll m_i$ one effectively has Dirac neutrinos.
This case requires $M_I< 10^{-9}$ eV, otherwise solar neutrino oscillations into $\nu_R$ should have been observed \cite{deGouvea:2009fp}, though LFV and cosmological constraints forbid that all $M_I$ are in this range \cite{}. 
Sterile neutrinos with $M_I$ in the eV range could also explain the ``oscillation anomalies'' (i.e. the LSND \cite{Athanassopoulos:1996jb}, Gallium \cite{Abdurashitov:2005tb} and reactor anomalies \cite{Mention:2011rk}) and/or act as extra relativistic degrees of freedom in the early universe ("dark radiation"). 
Note, however, that light sterile neutrinos that explain these anomalies cannot simultaneously explain the masses of active neutrinos; they would have to be added on top of the usual seesaw (\ref{activemass}). 
While all these anomalies as well as some cosmological data sets seem to favour  the existence of light sterile neutrinos, there is no convincing model that can fit all data sets simultaneously without significant tension.
On the other hand, the statistical significance is not sufficient to rule out their existence, even if the recent Planck data (which disfavours light sterile neutrinos) is taken into account, see Fig.~\ref{planck}. 
A more detailed discussion can e.g. be found in Ref.~\cite{Palazzo:2013me,KOPP:2014pua} (experimental side),  Ref.~\cite{Planck:2015xua} (most recent cosmological constraints) and in the reviews \cite{Abazajian:2012ys,Drewes:2013gca}.
Ultimately this question can only be clarified by new experiments.
\section{Conclusion}
To date, neutrino oscillations are the only established evidence for the existence of new physical states that has been found in the laboratory.
The recent years have seen immense progress in the determination of the neutrino  mixing angles and mass splittings. There are several important remaining questions about the properties of neutrinos that (with some luck) can be answered in foreseeable time, including the absolute mass scale, the nature of their mass term (Dirac vs Majorana) and the presence of CP-violation in the lepton sector.
The ultimate goal, however, remains to unveil the mechanism of neutrino mass generation, and to identify the new physical states that are involved in it. These could provide the key to understand other unsolved puzzles in both, cosmology and particle physics, as summarised in Appendix \ref{overview}. Unfortunately there is no guarantee that this can be achieved in foreseeable time.
In spite of this, it is important to explore the neutrino sector in as much detail as possible. 
Given the present lack of any new physics signals from high energy experiments, it remains the only probe of new physics that can be studied in the laboratory.

\appendix
\begin{landscape}
\section{Overview Table: Majorana mass scales and observables}\label{overview}
\setlength{\voffset}{20mm}
\scriptsize
\begin{center}
\begin{tabular}{| c || c | c | c | c | c | c | c |}
\hline
$M_M$ & Motivation  & $\upnu$-oscillations  & laboratory searches & indirect signals & BBN & DM & Leptogenesis \\
\hline\hline
$\lesssim$eV  & \begin{tabular}{c}$\upnu$-oscillations anomalies,\\ dark radiation\end{tabular} & 
\begin{tabular}{c}
\green{masses by seesaw,}$^a$\\
\green{explain anomalies}$^b$
\end{tabular}
& 
\begin{tabular}{c}
\green{oscillation anomalies,}\\
\green{$\beta$-decays}
\end{tabular} & \begin{tabular}{c}\green{CMB: explain $N_{\rm eff}>3$}$^b$ \\  LFV, $0\nu\beta\beta^g$\end{tabular} & \green{may explain $N_{\rm eff}>3$}$^b$ & \red{no} & \red{no} \\ 
\hline
keV  & DM & \red{no if DM}\green{$^c$} & 
\begin{tabular}{c}
\green{direct searches?}\red{$^d$}\black{,}\\
\green{$\beta$-decays}
\end{tabular}
& \begin{tabular}{c}\black{if DM:}\green{nuclear decays?}\red{$^d$}\black{,} \\ \green{pulsar kicks, supernovae}\\
\black{if not DM also \green{LFV, $0\nu\beta\beta^g$}}
\end{tabular} & \begin{tabular}{c} 
\red{effect on $N_{\rm eff}$}\\ \red{too small if DM}\end{tabular}  & \green{good candidate} & \red{no}  \\ 
\hline
MeV  & testability, why not? & \green{masses by seesaw}  & 
%\begin{tabular}{c}can be found,\\ constraints exist\end{tabular}  
\begin{tabular}{c}
\green{intensity frontier}
\end{tabular}
& \green{$0\nu\beta\beta$} & \begin{tabular}{c}\green{constrains}\\ \green{$M_I\gtrsim 100$ MeV}
%\\ (with direct search bounds)
\end{tabular} & \red{no}\green{$^e$}%\footnote{Unless $F\simeq0$ ensures stability and production is due to an unknown interaction.} 
& \begin{tabular}{c}\green{possible}\\ \green{(fine tuning)}\end{tabular} \\ 
\hline
GeV  & \begin{tabular}{c}testability,\\ minimality\end{tabular}   & \green{masses by seesaw} & \begin{tabular}{c}
\green{intensity frontier} 
\end{tabular} 
& \begin{tabular}{c}\green{EW precision data, LFV}\\ 
\green{$0\upnu\beta\beta$},\\
\green{lepton universality} \end{tabular}& \red{unaffected} & \red{no}\green{$^e$} & \green{possible} \\ 
\hline
TeV  & \begin{tabular}{c}minimality,\\testability\end{tabular}  & \green{masses by seesaw} & \green{LHC, FCC} & \begin{tabular}{c}\green{EW precision data,}\\ 
\green{$0\upnu\beta\beta$, LFV$^f$}\\
\green{lepton universality} \end{tabular} & \red{unaffected} & \red{no}\green{$^e$} & \green{possible} \\ 
\hline
$\gg$ TeV  & \begin{tabular}{c}grand unification,\\ ``naturally'' small $\upnu$-masses\end{tabular}  & \green{masses by seesaw} & \red{too heavy to be found} & \green{$0\upnu\beta\beta$, LFV$^{f}$}  & \red{unaffected} & \red{no}\green{$^e$} & \begin{tabular}{c}\green{works naturally}%for wide\\ range of parameters
\end{tabular}\\  
\hline
\end{tabular}
\end{center}
Colour code: \green{green = can affect}, 
%\yellow{may affect}, 
\red{red = does not affect}\\
\black{$^a$ At least one of the $N_I$ must be heavier than 100 MeV, see Fig.~\ref{allowedrange}. Even in that case, the eV-seesaw is under pressure from the combination of LFV, BBN constraints, CMB data and neutrino oscillation data \cite{Hernandez:2014fha}.}\\
\black{$^b$ Sterile neutrinos that may explain the oscillation anomalies cannot be the same ones as those that generate the observed active neutrino mass splittings  $\Delta m_{\rm atm}$ and $\Delta m_{\rm sol}$. 
Moreover, sterile neutrinos with masses and mixings suggested by the oscillation anomalies would be in thermal equilibrium in the early universe, hence increase $N_{\rm eff}$ by one unit per species. In $\Lambda$CDM cosmology, this is disfavoured by recent CMB data \cite{Planck:2015xua}. If some mechanism prevents the sterile neutrinos from getting into thermal equilibrium or there are deviations from the standard $\Lambda$CDM model, this conflict can be avoided.}\\
$^c$ Sterile neutrinos that compose the observed DM cannot give a sizable contribution to $m_\nu$ because their Yukawa couplings must be very small to suppress their decay.\\
$^d$ It is disputed whether the signal can be distinguished from the active neutrino background \cite{Ando:2010ye}; for the case that keV sterile neutrinos compose all DM, searches as proposed in Ref.~\cite{Li:2010vy} would be extremely challenging because of the astrophysical constraints on the mixing angle.\\
$^e$ This applies to sterile neutrinos thermally produced via their mixing. Sterile neutrinos with $M_I\gg$ keV can be DM if $F\simeq0$ ensures their stability and the production in the early universe is due to an unknown interaction.\\
$^f$ The rate is in general too small to be observed unless there is either an approximately conserved lepton number that allows for large Yukawa couplings or physics beyond the minimal seesaw is involved (such as supersymmetry \cite{Casas:2001sr}).\\
$^g$ If $M_M$ is the only source of lepton number violation, then the rates of neutrinoless double $\beta$-decay ($0\nu\beta\beta$) and other lepton number violating processes can only be large enough to be observed if at least one $M_I$ is larger than $100$ MeV. If this is given, then the lighter $N_I$ with $M_I<100$ MeV can contribute significantly to $0\nu\beta\beta$, see e.g. \cite{LopezPavon:2012zg,Hernandez:2014fha,Bilenky:2014uka}. 
\normalsize
\end{landscape}

\end{document}